\documentclass[
reprint,
superscriptaddress,
amsmath,amssymb,
aps,
prb,
]{revtex4-1}
\usepackage[utf8]{inputenc}
\usepackage{graphicx}
\usepackage{dcolumn}
\usepackage{bm}
\usepackage[normalem]{ulem}
\usepackage[colorlinks,citecolor=blue,linkcolor=blue,urlcolor=blue]{hyperref}
\usepackage{xcolor}

\begin{document}

\title{
Spin-orbital liquids and insulator-metal transitions on the pyrochlore lattice
}

\author{Nyayabanta Swain}
\affiliation{
Harish-Chandra Research Institute
(A CI of Homi Bhabha National Institute),
Chhatnag Road, Jhunsi, Prayagraj 211019, India}

\author{Madhuparna Karmakar}
\affiliation{Centre for Quantum Science and Technology, 
Chennai Institute of Technology, Chennai-600037, India.}

\author{Pinaki Majumdar}
\affiliation{
Harish-Chandra Research Institute
(A CI of Homi Bhabha National Institute),
Chhatnag Road, Jhunsi, Prayagraj 211019, India}

\date{\today}

\begin{abstract}
The two orbital Hubbard model, with the electrons additionally coupled to a 
complex magnetic background, arises in the pyrochlore molybdates. The  
background involves local moments Hund's coupled to the electrons, driving  
double exchange ferromagnetism, and antiferromagnetic (AF) tendency arising from 
competing superexchange.  The key scales include the Hubbard repulsion and 
the superexchange, both of which can be tuned in these materials. They control 
the phase transition from a ferromagnetic metal to a spin glass metal and then 
to a spin glass (Mott) insulator. We provide a comprehensive description of the 
ground state of this model using an unrestricted Hartree-Fock scheme implemented 
via a simulated annealing procedure and  establish the metal-insulator transition 
line for varying Hubbard interaction and superexchange. The electrons see an
effective disorder, due to orbital frustration, already in the ferromagnetic 
phase. The disorder is further enhanced by  antiferromagnetic coupling and the 
resulting magnetic disorder.  As a result, increasing AF coupling shifts the 
metal-insulator transition to lower Hubbard interaction and gives it an 
additional  ``Anderson'' character. We provide detailed results on the 
magnetic and orbital correlations, the density of states, and the optical 
conductivity. 
\end{abstract}


\maketitle

\section{Introduction}

The most commonly studied Mott problem \cite{MIT_Mott}
involves the single band Hubbard model on a bipartite lattice 
\cite{Mott_rev}. 
In such a model, typically, 
nesting features drive a transition \cite{FS_nesting_2d} 
to an antiferromagnetic insulating state at
arbitrarily weak interaction - masking the `Mott' effect.
One can certainly study frustrated lattices \cite{frust-latt}, 
which suppress magnetic order, and there is much work
on the triangular lattice Hubbard model\cite{tr_mod1,tr_mod2,tr_mod3,tr_mod4,tr_mod5}. 
This is both an important model problem \cite{SL-triangle}
and also a starting point for the layered organics
\cite{SL-tr-organic1,SL-tr-organic2,SL-tr-organic3,SL-tr-organic4,SL-tr-organic5}.
Other frustrated lattices include the Kagome 
\cite{frust-Kagome-lattice,Kagome-Hubbard-model,Kagome-MIT_CDMFT,Kagome-MIT_VCA,Kagome-MIT_CDMFT2,Kagome-MIT_DMRG,Kagome-MIT_DMFT_DQMC}
in two dimensions (2D) and the FCC and 
pyrochlore lattices \cite{pyr-rmp,pyr-recent-rev1,pyr-recent-rev2,pyr-recent-rev3} in 3D. 
These are all harder problems than the
square (or cubic) lattice since there is
no longer any obvious magnetic order to
simplify the correlated problem. These lattices,
overall, provide interesting variation from the
bipartite case because (i)~the metal-insulator
transition could occur in the background of short-range
magnetic correlation, 
and (ii)~the deep Mott 
insulating state itself could be a spin liquid 
\cite{QSL-frustration}. 

It would be vital to have experimental realisations 
to test out the predictions of the frustrated Mott
studies. While there is significant effort in
analysing the quasi 2D $\kappa$-BEDT organics
\cite{SL-tr-organic1,SL-tr-organic2,SL-tr-organic3,SL-tr-organic4,SL-tr-organic5} 
in terms of the triangular lattice, 
3D realisations of `Hubbard physics' 
on a frustrated structure are rare.
Materials like the manganites \cite{manganite} do involve
strong correlation effects (and much else) but
are on a bipartite structure - with relatively simple
magnetic order. The discovery of the
rare-earth (R) based pyrochlores, 
the molybdates \cite{pyr-Mo-sf1,pyr-Mo-sf2,Mo-Tc-Tf,
Mo-MIT-chem-pr,Mo-MIT-chem-pr2}, R$_2$Mo$_2$O$_7$,
and the iridates \cite{pyr-Ir-chem-pr,pyr-Ir-chem-pr2,
pyr-Ir-hydro-pr,pyr-Ir-hydro-pr2}, R$_2$Ir$_2$O$_7$, 
provided a breakthrough. 
Both these families show a metal-insulator transition as
the rare-earth radius $r_R$ is reduced
\cite{Mo-MIT-chem-pr,Mo-MIT-chem-pr2,pyr-Ir-chem-pr,pyr-Ir-chem-pr2}. 
There are, however, key differences between these
two families:
(i)~in terms of degrees of freedom and couplings, with
respect to the Hubbard model, and (ii)~the magnetic 
state that emerges.

Being 4d and 5d systems, respectively, both 
molybdates and iridates involve multiple bands.
In the molybdate case this can be reduced to 
one itinerant electron in two degenerate orbitals. 
These electrons have an inter-orbital Hubbard repulsion
and are also Hund's coupled to a $S=1/2$ 
local moment at each Mo site \cite{molybdate-DFT1}.
For the iridates one can motivate the use of an
effective single band model
which involve strong spin-orbit coupling 
in addition to the Hubbard interaction \cite{iridate_1_band,pyr-recent-rev2}.
While both families show a `Mott' transition, for the
molybdates this happens in a somewhat spin disordered
background, with no long range order in the insulating
state \cite{pyr-rmp,pyr-Mo-sf1,pyr-Mo-sf2},
while the iridates generally show a transition
from a paramagnetic metal to an `all-in-all-out' 
magnetic insulator 
\cite{iridate_AIAO_0,iridate_AIAO_1,iridate_AIAO_2}. 
The frustration in the pyrochlore
lattice plays a role in both these materials, but one
clearly requires more than the simple Hubbard model 
to approach the phenomena.

This paper is focused on a detailed study of the
model appropriate to the molybdates R$_2$Mo$_2$O$_7$.
These exhibit ground states 
that vary from a ferromagnetic metal (FM-M) 
to a spin glass metal (SG-M) and then a spin glass insulator 
(SG-I) as the rare earth radius is reduced
\cite{Mo-var-mag-ele1,Mo-var-mag-ele2}.
Materials with R = Nd and Sm are metallic,
R = Tb, Dy, Ho, Er, and Y are insulating, 
and R=Gd is on the verge of the insulator-metal transition
\cite{Mo-var-mag-ele2,Mo-optics2} (IMT).
The highest ferromagnetic $T_c$ is $\sim 100$K, in Nd, 
while the spin glass transition temperature, $T_{SG}$ is typically 
\cite{Mo-Tc-Tf1,Mo-Tc-Tf2,Mo-Tc-Tf3} $\sim 20$K.
The unusual features in transport include
very large residual resistivity, $\sim$ 10 $m\Omega$cm
close to the metal-insulator transition \cite{Mo-optics2},
prominent anomalous Hall effect in metallic samples
\cite{Mo-anm-hall1,Mo-anm-hall2,Mo-anm-hall3, Mo-anm-hall4,Mo-anm-hall5}, 
{\it e.g}, Nd$_2$Mo$_2$O$_7$,
and magnetic field driven metalization in the weakly insulating samples 
\cite{hanasaki-andmott} {\it e.g}, Gd$_2$Mo$_2$O$_7$.
Further, it's been shown that the local structural distortion
plays an important role in the stabilization of 
the ferromagnetic or spin-glass phases~\cite{CASTELLANO2017}.
More recent study of Mn-doping of Gd and Ho molybdates in the form
R$_2$(Mo$_{1-x}$Mn$_x$)$_2$O$_7$ demonstrate the strengthening of 
the spin-glass behavior in both systems upon doping~\cite{CASTELLANO2017}.
On the other hand, Ca-doping of Nd molybdate in the form
(R$_{1-x}$Ca$_x$)$_2$Mo$_2$O$_7$ with $x \leq 0.15$ displays 
robust canted ferromagnetic state. The stability of the canting 
angle of the Mo magnetic moments with respect to the doping is 
a key puzzle in the topological Hall effect observed in 
this material ~\cite{Ca-doping_Nd-molyb_2021, topo-Hall_Nd-molyb_2021}. 

We will discuss the model for the molybdates in detail later,
to motivate our study it suffices to mention that 
the active degrees of freedom include one electron per Mo
in a twofold degenerate orbital, Hund's coupled to a $S=1/2$
moment on the same ion. The electrons have onsite Hubbard
repulsion $(U)$ between them while the local moments have a nearest
neighbour antiferromagnetic (AF) coupling, $J_{AF}$.
The Hund's coupling drives 
double exchange (DE) ferromagnetism, opposed
by AF superexchange, while Hubbard repulsion promotes a Mott
insulating state. Reducing $r_R$ reduces the hopping - weakening
DE and also enhancing the effect of Hubbard repulsion, while
increasing pressure is supposed to (mainly) affect 
\cite{Mo-MIT-chem-pr2} the antiferromagnetic coupling.

There are several major questions left unresolved by
existing work:
(1)~At ambient pressure the metal-insulator and magnetic `transition'
are simultaneous, is that true with increasing pressure (changing 
$J_{AF}$) as well? (2)~Is there an `universal' quantity that 
dictates  the metal-insulator transition (MIT) 
trajectory over a large pressure window?
(3)~What is the fate of the coupled spin-orbital state
for changing pressure and rare earth radius?
(4)~What is the low energy spectral behaviour in the vicinity
of the MIT as the pressure is varied?
(5)~What is the quasiparticle character close to the 
Mott transition?
(6)~Can we obtain realistic thermal scales for the 
magnetic transitions?

We employ a real space approach, equivalent to unrestricted 
Hartree-Fock
at zero temperature, that uses a static auxiliary orbital
field to handle the Hubbard interaction. We solve the
resulting `electron~-~local moment~-~orbital moment' problem 
via Monte Carlo based simulated annealing 
on the pyrochlore lattice. Within the limits of our method
we address (1)-(4) of the questions posed above, and (5) and (6) elsewhere. 
Our main results are the following:

\begin{itemize}
\item {\it Phase boundaries:}
The proximity of the magnetic transition and MIT in the ambient 
pressure molybdates is a coincidence -  at weak AF coupling the
 metal and insulator are both ferromagnetic, while at strong AF 
 coupling they are both spin disordered.
\item {\it Physics behind the MIT:}
The shift in the critical interaction for the MIT, with applied pressure, 
can be understood in terms of the kinetic energy suppression driven 
by growing  spin and orbital disorder.
\item {\it Coupled spin-orbital state:}
The magnetic state is a spin ferromagnet (S-F) or a spin liquid (S-L), 
the orbital state is similarly 
orbital ferromagnet (O-F) or orbital liquid (O-L). We find that the 
low $J_{AF}$ state is mainly S-F~-~O-F while the large $J_{AF}$ 
state is S-L~-O-L.
\item {\it Spectral behaviour near the MIT:}
The $U_c$ changes with changing $J_{AF}$, so we use a normalised
 frequency scale, $\omega/U_c(J_{AF})$, to compare spectral features.
  At the MIT the larger $J_{AF}$ systems have more low energy spectral 
  weight than the weak $J_{AF}$ case. Surprisingly, the gap edge states 
  at large $J_{AF}$ are strongly localized, leading to an optical gap that
   is larger than the density of states gap, revealing the growing 
   Anderson character of the transition with increasing $J_{AF}$.
\end{itemize}

The remaining of this paper is structured as follows.
In section II we discuss the static auxiliary field based Monte Carlo 
method in connection with  the full fledged determinant quantum 
Monte Carlo and the unrestricted Hartree-Fock methods. 
This is followed by a discussion of our results in section III, comprising 
of the ground state phase diagram, the detailed magnetic and orbital 
structure factors in the different phases, and the density of states 
and transport properties across the
metal-insulator transition. We propose an effective spin only model 
for the fermionic system under consideration in section IV and discuss 
the MIT in the light of this effective model. We conclude in 
section V with pointers for experiments. 

\section{Model and method}

\subsection{Model}

The R$_2$Mo$_2$O$_7$ structure consists of two interpenetrating
pyrochlore lattices, one formed by Mo cations and the other by R.
Model Hamiltonian studies ignore the orbitals on R and oxygen,
focusing instead on the orbitals on Mo.
The Mo atom has octahedral oxygen coordination, the
resulting crystal field splits the fivefold
degenerate Mo 4d states into doubly degenerate $e_g$
and  triply degenerate $t_{2g}$ manifolds,
and a trigonal distortion splits the $t_{2g}$ further into a 
nondegenerate $a_{1g}$ and a doubly degenerate $e'_g$.
The hopping between Mo orbitals at different 
sites is mediated by the intervening oxygen.  The Mo cation 
is nominally tetravalent and has two electrons on average.
The deeper  $a_{1g}$ state behaves like a local moment,
and the single electron in the two  $e'_g$ orbitals
is the `itinerant' degree of freedom \cite{molybdate-DFT1}.
The $e_g$ state remains unoccupied.

There are additional small scales, related to bond distortions,
{\it etc}, that are responsible for the spin freezing phenomena
\cite{pyr_HAF_SG,Molyb-els-coup-th}.
We ignore them for the time being. Also,
the moments on R can be relevant when studying effects
like spin chirality induced anomalous Hall effect
\cite{Mo-anm-hall1,Mo-anm-hall2,Mo-anm-hall3, Mo-anm-hall4,Mo-anm-hall5}.
We do not include these moments in our model.

We study the following model \cite{Molyb-model}, 
in parameter regimes described below:
\begin{eqnarray}
H &= & \sum_{\langle ij \rangle,\alpha\beta,\sigma}
t_{ij}^{\alpha\beta} 
c^{\dagger}_{i\alpha\sigma}c_{j\beta\sigma}  
- J_H \sum_{i,\alpha} {\bf S}_i \cdot c^{\dagger}_{i\alpha\sigma}  
\vec{\sigma}_{\sigma\sigma^{'}} c_{i\alpha\sigma^{'}} \cr
&&~~ 
+ J_{AF} \sum_{\langle ij \rangle} {\bf S}_i.{\bf S}_{j} 
+ \sum_{i,\alpha\beta\alpha^{'}\beta^{'}}^{\sigma, \sigma'}
 U_{\alpha\beta}^{\alpha^{'}\beta^{'}}
 	c^{\dagger}_{i\alpha\sigma}c^{\dagger}_{i\beta\sigma^{'}}
c_{i\beta\sigma^{'}}c_{i\alpha\sigma} \nonumber  
\end{eqnarray}
The first term is the kinetic energy, involving nearest neighbour 
intra and inter-orbital $e'_g$ hopping.
The second term is the Hund's coupling 
between the $a_{1g}$ local moment ${\bf S}_i$ and the $e'_g$ 
electrons,
$J_{AF}$ is the AF superexchange coupling between local moments at
neighbouring sites on the pyrochlore lattice, and the $U$ 
represent onsite $e'_g$ Coulomb matrix elements.

To simplify the computational problem 
we treat the localized spins ${\bf S}_i$ 
as classical unit vectors, absorbing the size $S$ in
the magnetic couplings. We will comment on the 
limitations of this approximation later.
Also, to reduce the size of the Hilbert space
we assume that $J_H/t \gg 1$, where $t$ is the
typical hopping scale, so that only the
locally `spin aligned' fermion state is retained.
In this local basis the hopping matrix elements 
are dictated by the orientation of the ${\bf S}_i$ 
on neighbouring sites. 
This leads to the simpler model:
$$
H  =  \sum_{\langle ij \rangle,\alpha\beta}
{\tilde t}_{ij}^{\alpha\beta} {\tilde c}^{\dagger}_{i\alpha} {\tilde c}_{j\beta} 
   + J_{AF} \sum_{\langle ij \rangle} {\bf S}_i.{\bf S}_j
+ U \sum_{i}^{\alpha \neq \beta} n_{i\alpha} n_{i\beta} 
$$
where the fermions are now `spinless'.
$U >0$ is the inter-orbital Hubbard repulsion. 
The effective hopping is determined by the orientation of
the localized spins ${\bf S}_i = (sin\theta_{i}
cos\phi_{i},sin\theta_{i}sin\phi_{i},cos\theta_{i})$, as 
$t_{ij}^{\alpha\beta} = [ cos\frac{\theta_i}{2}
cos\frac{\theta_j}{2}
		 + sin\frac{\theta_i}{2}sin\frac{\theta_j}{2} 
e^{-i(\phi_i - \phi_j)}] t^{\alpha\beta} $,
with $t^{11} = t^{22} =t$ and $t^{12} = t^{21} =t'$.
We set $ t' = 1.5t$ as is appropriate for these kinds of
orbitals \cite{molybdate-DFT1}.

The first two terms represent fermions in a classical spin
background and the resulting magnetic phase competition has
been studied on a pyrochlore \cite{Molyb-DE-SE}. 
While these results are interesting 
they miss out on the large correlation scale,
$U$, that drives the Mott transition. One option
is to treat the model within dynamical mean field theory (DMFT)
\cite{DMFT-georges}, but then the spatial character 
crucial to the pyrochlore lattice is lost. 

The current paper is focused on the ground state but
we discuss our general strategy for solving the
finite temperature problem below. This will describe the 
simulated annealing scheme for arriving at the ground state.

\subsection{Method}

We handle the problem in real space
as follows: (i)~We use a Hubbard-Stratonovich (HS)
\cite{hubb-strat,hubbard,schulz} transformation 
that decouples $U n_{i \alpha} n_{i \beta}$ in terms of
an auxiliary orbital variable ${\bf \Gamma}_i(\tau)$,
coupling to the electronic orbital moment 
${\bf O}_i = \sum_{\mu \nu} c^{\dagger}_{i \mu}
{\vec \sigma}_{\mu  \nu} c_{i \nu}$, and 
a scalar field $\Phi_i(\tau)$ coupling to the electronic density 
$n_i$ at each site.
The Matsubara frequency versions of these fields are
${\bf \Gamma}_{i,n}$ and $\Phi_{i,n}$, where $\Omega_n = 2 \pi n T$ is
a bosonic frequency.
(ii)~An exact treatment of the resulting functional integral,
see below, requires determinant quantum
Monte Carlo (DQMC) - computing a fermion determinant 
$D({\bf \Gamma}_{i,n}, 
\Phi_{i,n}, {\bf S}_i)$
iteratively as the `weight factor' for auxiliary field 
configurations. Fermion Green's functions would be 
computed on the
equilibrium $\{ {\bf \Gamma}, \Phi, {\bf S}\}$ backgrounds.

The DQMC implementation, which we will approximate,
takes the following route.
The partition function 
is written as a functional integral 
over Grassmann fields $\psi_{i \alpha}(\tau)$ 
and ${\bar \psi}_{i \alpha}(\tau)$:
\begin{eqnarray}
Z & = &  \int {\cal D} \psi {\cal D} 
{\bar \psi} 
{\cal D} {\bf S}
e^{ -\int_{0}^{\beta}d \tau {\cal L}(\tau)} \cr
{\cal L}(\tau) & = & 
\sum_{\langle ij \rangle ,\alpha \beta} \{ {\bar \psi}_{i\alpha} 
((\partial_{\tau} - \mu) \delta_{ij}\delta_{\alpha \beta} 
+ {\tilde t}_{ij}^{\alpha \beta}) \psi_{j\beta} \} \cr
&& ~~~+ U \sum_{i ,\alpha \neq \beta} {\bar \psi}_{i\alpha} 
{\psi}_{i\alpha} {\bar \psi}_{i\beta} {\psi}_{i\beta}
~+~J_{AF} \sum_{\langle ij \rangle} {\bf S}_i.{\bf S}_j
\nonumber
\end{eqnarray}

The quartic term is `decoupled' exactly 
via a Hubbard-Stratonovich transformation
$$
e^{U {\bar \psi}_{i\alpha} {\psi}_{i\alpha} 
{\bar \psi}_{i\beta} {\psi}_{i\beta}} 
= \int \frac{d\Phi_i d{\bf \Gamma}_i}{4\pi^2U}
  e^{ ( i\Phi_i n_i - {\bf \Gamma}_i.{\bf O}_i
       + \frac{\Phi_i^2}{U} + \frac{\Gamma_i^2}{U}) }
$$
where $\Phi_{i}(\tau)$ and ${\bf \Gamma}_i(\tau)$ 
are two auxiliary fields: 
$\Phi_{i}(\tau)$ coupling to charge density 
$n_i =n_{i\alpha}+n_{i\beta}$, and
${\bf \Gamma}_i(\tau)$ coupling to the
orbital variable
${\bf O}_i = \sum_{\mu \nu} {\bar \psi}_{i \mu}
{\vec \sigma}_{\mu  \nu} {\psi}_{i \nu}$.
This leads to:
\begin{eqnarray}
Z~~ &=& \int {\cal D} \psi {\cal D} {\bar \psi} {\cal D} {\bf S}
\prod_{i}\frac{d\Phi_{i} d{\bf \Gamma}_{i}}{4\pi^{2}U}
      e^{-\int_{0}^{\beta} d \tau {\cal L}(\tau)} \cr
~~\cr
{\cal L}(\tau) & = &  {\cal L}_{0}(\tau) + {\cal L}_{int}(\tau) + 
{\cal L}_{cl}(\tau) \cr
~~\cr
{\cal L}_{0}(\tau) & = & \sum_{\langle ij \rangle ,\alpha\beta}
\{ {\bar \psi}_{i\alpha} 
( (\partial_{\tau} - \mu) \delta_{ij}\delta_{\alpha\beta} + 
{\tilde t}_{ij}^{\alpha\beta}) \psi_{j\nu} \} \cr
{\cal L}_{int}(\tau) & = & ~\sum_{i}^{\alpha \beta}
\{i\Phi_{i}{\bar \psi}_{i\alpha} \psi_{i\beta} \delta_{\alpha \beta}
- {\bf \Gamma}_{i}.{\bar \psi}_{i\alpha}\vec{\sigma}_{i}\psi_{i\beta} \}  \cr
{\cal L}_{cl}(\tau) & = & ~\sum_{i}\{\frac{\Phi_{i}^2}{U} +
 \frac{{\bf \Gamma}_i^2}{U} \}  
+ J_{AF} \sum_{\langle ij \rangle} {\bf S}_i.{\bf S}_j
\nonumber 
\end{eqnarray}
Since the fermions are now quadratic the $\int {\cal D} \Psi ..$
integrals can be formally performed to generate the effective
action for the background fields:
\begin{eqnarray}
Z~~ & \sim & \int {\cal D}\Phi {\cal D} {\bf \Gamma} {\cal D} {\bf S}
      e^{-S_{eff}\{ \Phi, {\bf \Gamma}, {\bf S} \}} 
\cr
S_{eff} &=& log~Det[{\cal G}^{-1}\{ \Phi, {\bf \Gamma}, {\bf S} \}]
+ \int_0^\beta d \tau {\cal L}_{cl}(\tau) 
\nonumber
\end{eqnarray}
In the expression above ${\cal G}$ is the electron Green's function in a 
$\{ \Phi, {\bf \Gamma}, {\bf S} \}$ background.

Now the options.
(1)~Determinant Quantum Monte Carlo would proceed by using 
$S_{eff}$ as the `weight'
for the background configurations, and compute  electron properties
on these after equilibriation.
(2)~Mean field theory would assume the fields to be time independent,
replace them by their mean values, and minimise the free energy.
(3)~A {\it static path approximation} to $Z$ again assumes the fields
to be time independent, but samples over spatial fluctuations.

We adopt method (3), which is computationally simpler than
DQMC but much more sophisticated than MFT at finite temperature.
So we
(i)~neglect the imaginary 
time dependence of $\Phi_i$ and ${\bf \Gamma}_{i}$, 
{\it i.e}, retain only the zero Matsubara frequency modes of these
fields, and 
(ii)~replace $\Phi_i$ by its saddle point value 
$\langle \Phi_i \rangle = (U/2)\langle n_i \rangle = U/2$,
since the important low energy fluctuations arise from the
 ${\bf \Gamma}_i$. 
The electron is now subject to {\it static} background fields
so the partition function can be written as a trace over an 
effective `Hamiltonian', rather than require an effective `action'.
Specifically:
\begin{eqnarray}
H_{eff}\{{\bf \Gamma}_i, {\bf S}_i\} &=& 
-\frac{1}{\beta} logTr e^{-\beta H_{el}} 
+ H_{AF} 
+ \frac{1}{U} \sum_{i} {\bf \Gamma}_i^2
\cr
H_{el} & = & \sum_{ij}^{\alpha\beta} {\tilde t}_{ij}^{\alpha\beta} 
c^{\dagger}_{i\alpha}c_{j\beta}  - {\tilde \mu} \sum_i n_i 
- \sum_{i} {\bf \Gamma}_i.{\bf O}_{i} 
\nonumber
\end{eqnarray}
with ${\tilde \mu} = \mu - U/2$ and $H_{AF}$ the Heisenberg term.
For convenience we redefine 
${\bf \Gamma}_i  \rightarrow \frac{U}{2} {\bf \Gamma}_i $, 
so that the ${\bf \Gamma}_i$ is dimensionless.
This leads to the effective electronic 
Hamiltonian used in the text:
$$
H_{el} = \sum_{ij}^{\alpha\beta} {\tilde t}_{ij}^{\alpha\beta} 
c^{\dagger}_{i\alpha}c_{j\beta} - {\tilde \mu} \sum_i n_i 
- {U \over 2} \sum_{i} {\bf \Gamma}_i.{\bf O}_{i}
$$
The localized spin and orbital moment configurations  
follow the distribution
$$
P\{{\bf S}_i, {\bf \Gamma}_i\} \propto 
\textrm{Tr}_{cc^{\dagger}} e^{-\beta H_{eff} }
$$
This overall approach has been used in the nuclear many body
problem \cite{SPA_nuclear_pf,SPA_nuclear_ld}, superconductivity
\cite{SPA_dagotto,SPA_meir,SPA_fflo,SPA_lieb}, {\it etc}, and by us in 
other studies of the Mott problem before \cite{mott-Tr,mott-pyr}.

There are regimes where some analytic progress can be made,
as we discuss later, but our results are based on a Monte Carlo
solution of the model above - generating the equilibrium
configuration for the $\{ {\bf S}_i, {\bf \Gamma}_i\}$
through iterative diagonalisation of $H_{eff}$. We start
with high temperature, $\sim 0.5t$, 
higher than any transition temperature
in the problem, and reduce it to $T = 0.001t$ to
access ground state properties. 
To access large sizes within reasonable time we
use a cluster algorithm \cite{tca}
for estimating the update cost.
Results in this paper are for a $6 \times 6 \times 6$
pyrochlore lattice of $\sim 800$ atoms.

A couple of comments on the $T \rightarrow 0$ limit of our method
which reduces to unrestricted Hartree-Fock in the magnetic channel.
Traditionally, Hartree-Fock calculations impose a certain pattern
on the order parameter and minimise with respect to the amplitude.
On a frustrated geometry it is not clear what pattern to impose
so we vary with respect to the full set $\{{\bf S}_i, {\bf \Gamma}_i\}$.
The resulting state turns out to be disordered but correlated, and
leads to a non trivial electronic spectrum.

\subsection{Observables}

From the equilibrium configurations obtained at the
end of annealing we 
calculate the following averaged quantities
(angular brackets represent thermal average over MC
configurations): 
(i)~Magnetic and orbital structure factors are: 
\begin{eqnarray}
S_{mag}({\bf q}) &=& 
\frac{1}{N^2}\sum_{ij}\langle{\bf S}_i.{\bf S}_j
\rangle e^{i{\bf q}\cdot({\bf r}_i-{\bf r}_j)} \cr
S_{orb}({\bf q}) & =& 
\frac{1}{N^2}\sum_{ij}\langle{\bf \Gamma}_i.{\bf \Gamma}_j
\rangle e^{i{\bf q}\cdot({\bf r}_i-{\bf r}_j)}
\nonumber
\end{eqnarray}
(ii)~The size distribution of the orbital field is computed
as
$$
P(\Gamma) = {1 \over N} \sum_i \langle \delta(\Gamma -
\vert {\bf \Gamma}_i \vert) \rangle
$$
(iii)~The electronic density of states is,
$$
N(\omega) = {1 \over N} \sum_n \langle
\delta(\omega - \epsilon_n) \rangle
$$
where $\epsilon_n$ are single particle eigenvalues 
in an equilibrium configuration.

(iv)~The optical conductivity is:
\begin{eqnarray}
\sigma_{xx}(\omega)&=&
\frac{\sigma_{0}}{N}
\langle \sum_{n,m}
{ {f(\epsilon_{n})- f(\epsilon_{m})}
\over {\epsilon_{m}-\epsilon_{n}} }
|J^{nm}_x|^2 \delta(\omega-E_{mn})
\rangle
\nonumber
\end{eqnarray}
where $J^{nm}_x$ is $\langle n \vert J_x \vert m \rangle$
and the current operator is given by
\begin{eqnarray}
J_x&=&-i\sum_{i,\alpha\beta}\left[({\tilde t}_{i,i+\hat{x}}^{\alpha\beta}
c^{\dagger}_{i,\alpha}c_{i+\hat{x},\beta}-\textrm{hc})\right]
\nonumber
\end{eqnarray}
$E_{mn} = \epsilon_{m}-\epsilon_{n}$,
$f(\epsilon_{n})$ is the Fermi function,
$\epsilon_{n}$ and $|n\rangle$ are
the single particle eigenvalues and eigenstates of
$H_{el}$ respectively.
The conductivity is in units of
$\sigma_{0} = e^2/(\hbar a_0)$,
where $a_0$ is the lattice constant.
(v)~The d.c. conductivity is obtained as a low frequency average of the
optical conductivity over a window $\Omega = 0.05t$.
$$
\sigma_{dc} = {1 \over \Omega} \int_0^{\Omega}
d \omega \sigma_{xx}(\omega)
$$
and the resistivity $\rho = 1/\sigma_{dc}$.

\begin{figure}[t]
\centerline{
\includegraphics[angle=0,width=4.5cm,height=4.8cm]{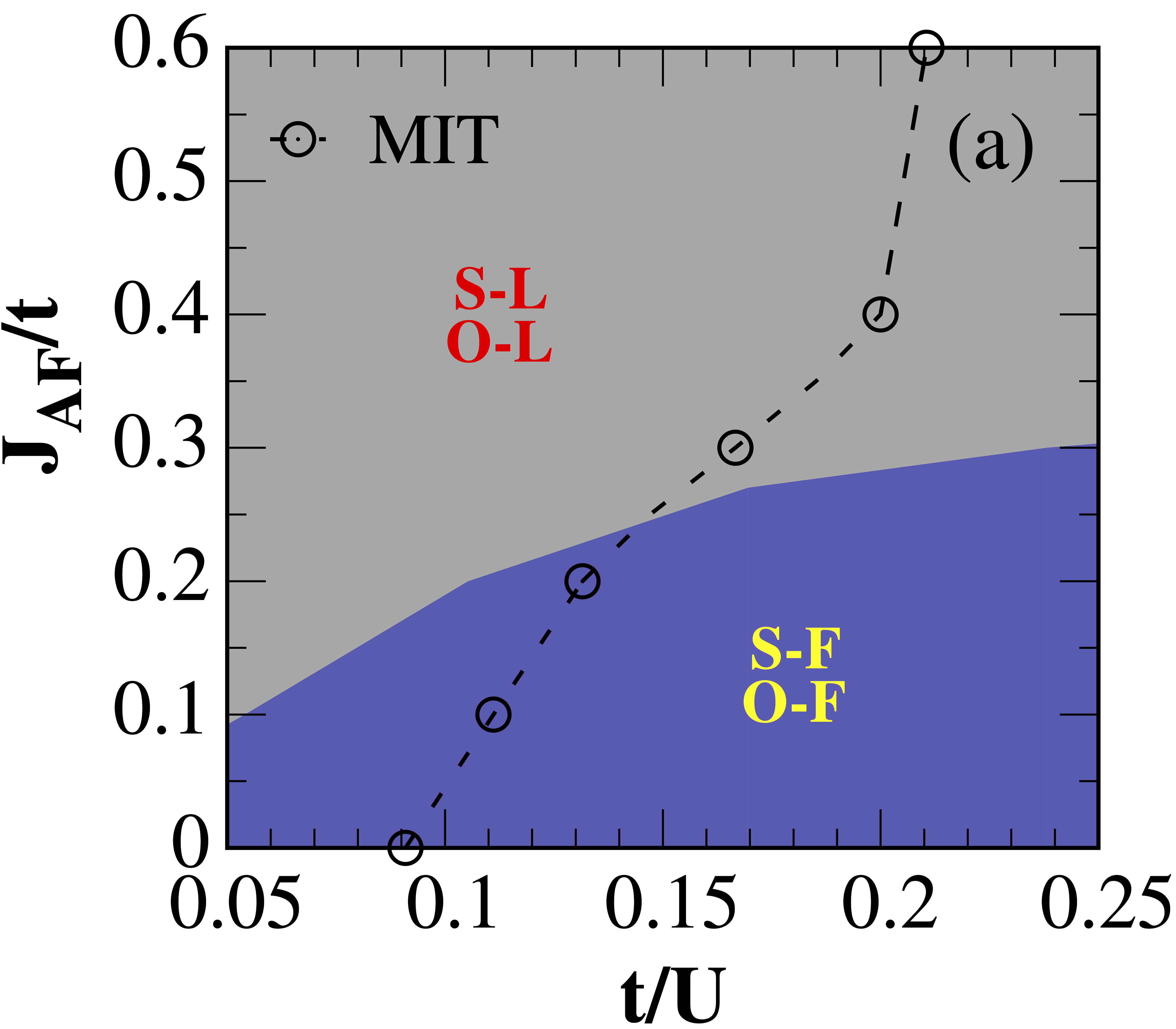}
\hspace{-0.15cm}
\includegraphics[angle=0,width=4.3cm,height=4.8cm]{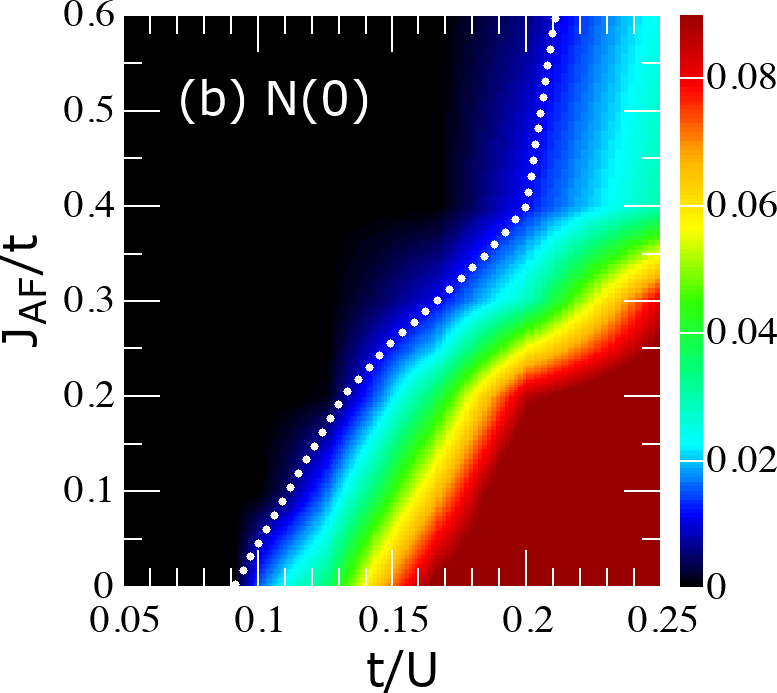}
}
\caption{\label{pd_T0} 
(a)~Ground state phase diagram showing 
the metal-insulator transition (MIT) boundary 
in the $t/U$, and $J_{AF}/t$ plane. 
We label the various magnetic phases  as 
spin-ferromagnet (S-F) and spin liquid (S-L).
The two orbital phases are labeled as 
orbital-ferromagnet (O-F) and orbital liquid (O-L).
The detailed chacterisation of these phases is
 mentioned in the text. 
Panel (b) shows the density of states at the Fermi level,
 $N(0)$, for varying $t/U$ and $J_{AF}/t$. 
The dotted line corresponds to MIT. $N(0)$ vanishes in the insulating 
phase at weak $J_{AF}/t$. However, it retains a small non-zero 
value at large $J_{AF}/t$, an indication of frustration induced 
Anderson localization behavior (discussed later).     
}
\end{figure}

\section{Results}

\subsection{Phase diagram}

Fig.\ref{pd_T0}(a) shows the ground state of the model for
varying $U/t$ and $J_{AF}/t$, while Fig.\ref{pd_T0}(b) shows
$N(0)$, the density of states at the Fermi level, over the same
parameter space. 

First the notation:
we characterise phases in terms of their spin and orbital
character, S-L is spin-liquid and S-F is
a spin ferromagnet. 
Similarly, O-L is orbital-liquid, {\it etc}. 
These phases also need to
be specified in terms of their transport character. To avoid
a cluttered picture we have simply shown the metal-insulator 
boundary in the $t/U-J_{AF}/t$ plane, the metal/insulator aspect
can be inferred from it. The metal-insulator transition can be
located from the vanishing of $N(0)$, and also from a calculation
of the d.c conductivity.

\begin{figure*}[t]
\centerline{
\includegraphics[angle=0,width=12.5cm,height=13.5cm]{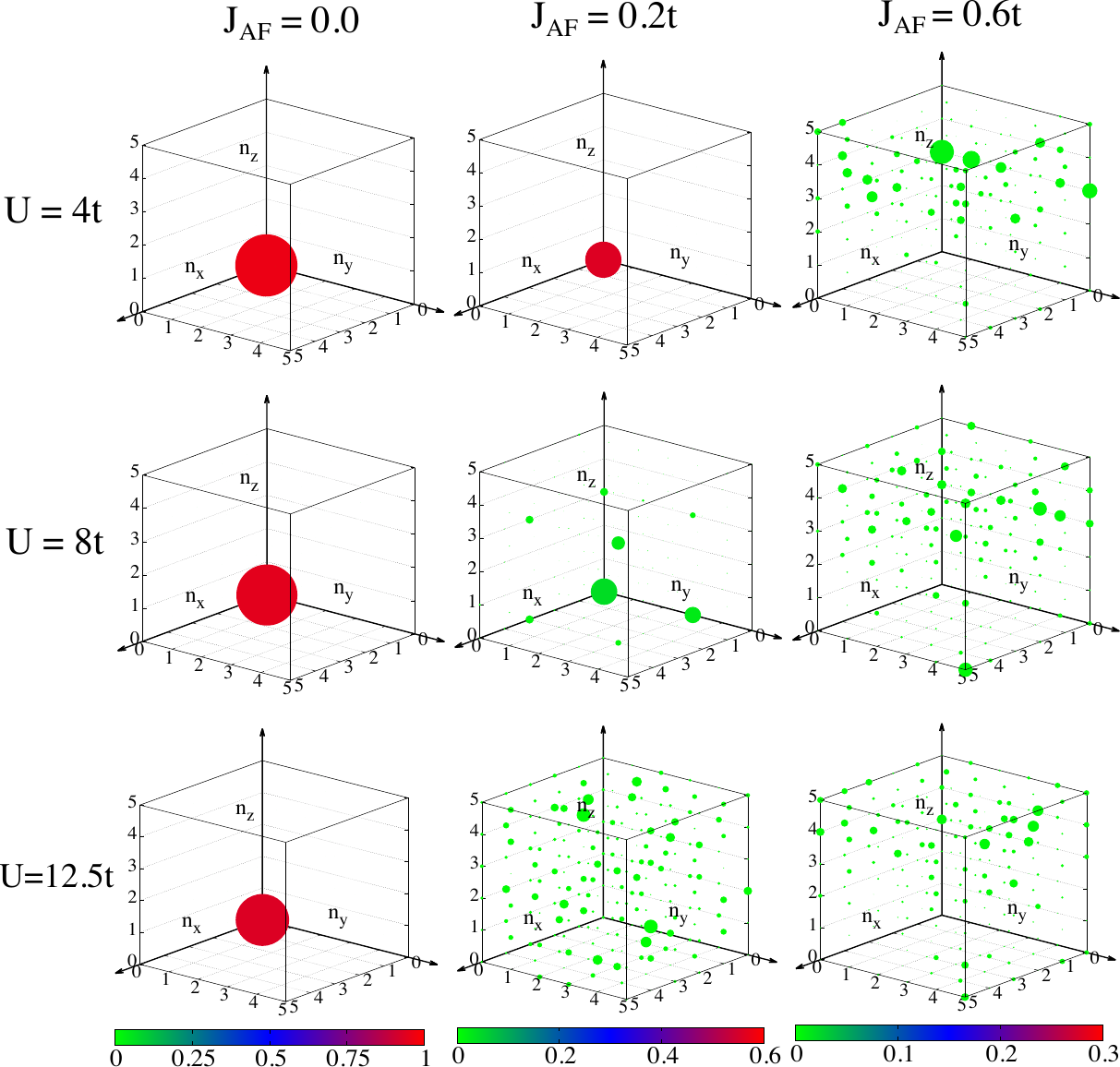}
}
\caption{\label{spin}
Spin structure factor $S_{mag}({\bf q})$ at $T=0$
for $U/t = $4, 8 and 12.5 for each of $J_{AF}/t = $0 (left column),
0.2 (middle column) and 0.6 (right column).
We use the notation ${\bf q} = \frac{2\pi}{L}(n_x,n_y,n_z)$, 
where $n_i$'s are integers and $0 \leq n_i < L$. In our calculation $L = 6$.
The  size of a dot signifies relative weight at a given
${\bf q}$ while its color represents the actual magnitude
of $S_{mag}({\bf q})$.
The presence of dominant weight at some ${\bf q}$, in these cases
${\bf q} = (0,0,0)$
indicates magnetic order phase, while the `random' but correlated
patterns indicate a spin liquid.
}
\end{figure*}

When $J_{AF} =0$ there is a
metal-insulator transition at $U_c \sim 11t$ from a 
ferromagnetic metal to a {\it ferromagnetic insulator}.
When the superexchange is moderate, $J_{AF} \sim 0.2t$,
there is strong competition between ferromagnetism
(S-F, mediated by double-exchange) and antiferromagnetic
tendency. As a result 
there is a crossover from S-F to spin disordered
(S-L) behaviour with increasing $U/t$ roughly around the MIT,
although weak ferromagnetism survives in the insulator.
For strong superexchange, $J_{AF} \gtrsim 0.5t$,
the antiferromagnetic tendency suppresses ferromagnetism
completely and, as we will show, there is no
magnetisation at any $U/t$.
We have a spin liquid (S-L) state at all $U/t$.
In this large $J_{AF}$ limit, 
a relatively weak Hubbard repulsion, $U \sim 5t$,
is enough to drive the metal-insulator transition. 

\begin{figure*}[t]
\centerline{
\includegraphics[angle=0,width=12.5cm,height=4.0cm]{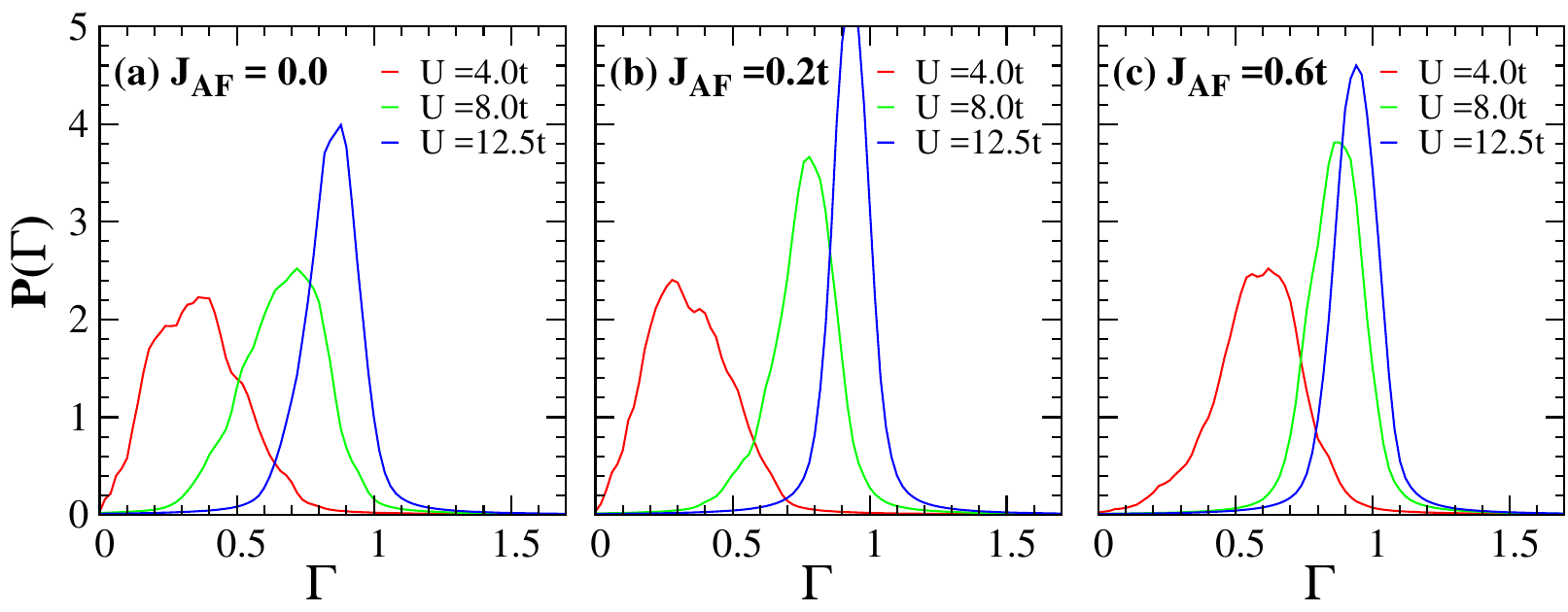}
~~~~~~~
}
\vspace{.3cm}
\centerline{
\includegraphics[angle=0,width=12.5cm,height=13.5cm]{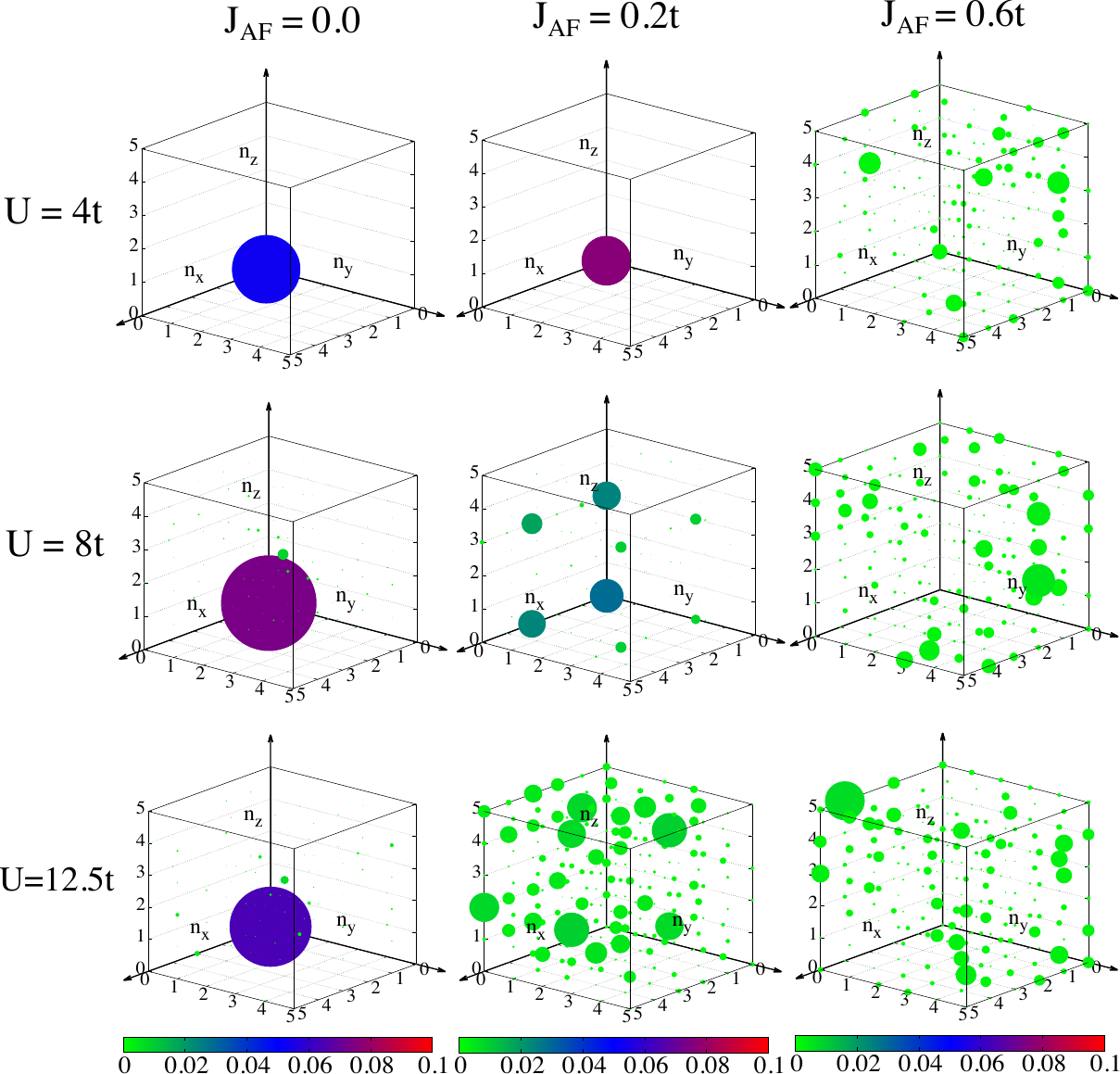}
}
\caption{\label{orb}
(Top row) Ground state size distribution of the orbital field $P(\Gamma)$
for $J_{AF} = $ 0, 0.2t and 0.6t for indicated $U$ values.  (Remaining rows) 
Orbital structure factor at $T=0$ for $U/t = $4, 8 and 12.5 for each of 
$J_{AF}/t = $ 0 (left column), 0.2 (middle column) and 0.6 (right column).
We use the same convention as described in Fig. \ref{spin} .  The  size of 
a dot signifies relative weight at a given ${\bf q}$ while its color 
represents the actual magnitude of $S_{orb}({\bf q})$.  The presence of 
dominant weight at some ${\bf q}$, indicates an orbital ordered phase, 
otherwise a disordered phase.
}
\end{figure*}

\begin{figure*}[t]
\centerline{
\includegraphics[angle=0,width=16.2cm,height=8.4cm]{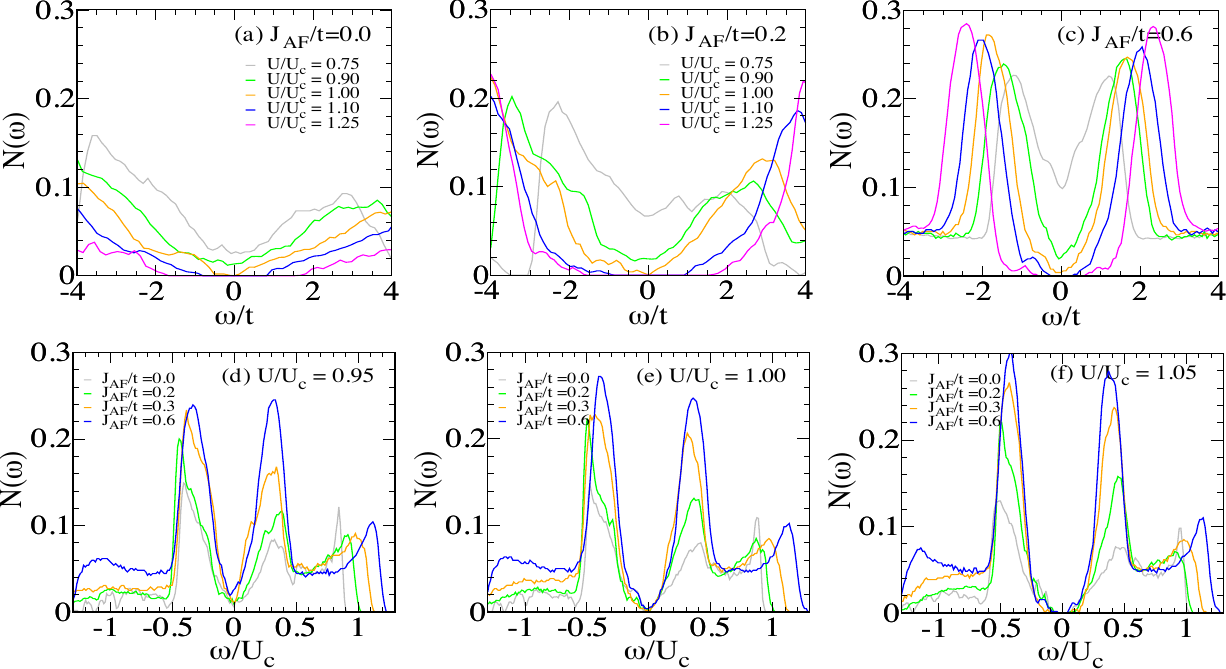}
}
\caption{\label{dos}
(a)-(c) Ground state density of states (DOS) for $J_{AF} =$ 0, 0.2t and 0.6t
for different $U/U_c$.
(d)-(f) Ground state DOS for $U/U_c =$ 0.95, 1.0 and 1.05,
on a normalised frequency scale, for the indicated $J_{AF}$ values. 
}
\end{figure*}

\subsection{The magnetic state}

A detailed understanding of  the magnetic state is
provided by the magnetic structure factor
$S_{mag}({\bf q})$
computed in the optimised background. It highlights not
only long range order, in terms of prominent peaks in
${\bf q}$ space, but also possible correlations in
the disordered state when there is no long range order. 

Fig.\ref{spin} shows $S_{mag}({\bf q})$ for three different 
superexchange couplings and for three $U$'s in each case.
The $U$'s are chosen so that they capture the metal, insulator,
and crossover regime for all three values of $J_{AF}$.

For $J_{AF}=0$
there is no magnetic phase competition.  At 
$U=4t$, $S_{mag}({\bf q})$ has dominant weight
at ${\bf q} =(0,0,0)$ describing the ferromagnetic order
promoted by double-exchange.
The magnetisation is $\gtrsim 0.95$ (limited by our annealing
process) and the structure factor peak is $\sim 0.9 \sim M^2$.
As the column shows, this result does not depend on $U$, 
suggesting that even deep in the Mott insulator 
one would obtain a saturated ferromagnetic state. 
The $T_c$'s would of course differ, since the stiffness of the
FM state depends on the kinetic energy - which is $U$
dependent.

For $J_{AF}=0.2t$, $S_{mag}({\bf q})$ has a large weight
at ${\bf q} =(0,0,0)$ at $U=4t$, as in the first row, but
at $U=8t$ the peak, although still at $(0,0,0)$, has
diminished weight, $\sim 0.1$.
The metal-insulator transition occurs around $U \sim 8t$ and
by the time $U=12.5t$ (last row) $S_{mag}$ does
not have any prominent peaks at any ${\bf q}$.
The superexchange coupling overcomes the kinetic
energy gain from DE but the pyrochlore structure
prevents AF ordering. 

For $J_{AF}=0.6t$, $S_{mag}({\bf q})$ the weight is spread over
all ${\bf q}$ but in a correlated manner, 
indicative of a spin liquid phase.

\subsection{The orbital state}

To have an idea of the underlying orbital state, 
we calculate the orbital structure factor 
$S_{orb}({\bf q})$.
Fig.\ref{orb} shows the structure factor
for the three superexchange couplings.
For $J_{AF}=0$ 
we see $S_{orb}({\bf q})$ has dominant weight
at ${\bf q} =(0,0,0)$ describing the orbital-ferro (O-F) ordering.
For $J_{AF}=0.2t$, $S_{orb}({\bf q})$ has dominant weight
at ${\bf q} =(0,0,0)$ for $U=4t$ and $8t$ (O-F ordering),
and an orbital liquid state for $U=12.5t$.
For $J_{AF}=0.6t$, $S_{orb}({\bf q})$ has weight
spread over all ${\bf q}$ indicating an orbital liquid state.

\subsection{Density of states}

Fig.\ref{dos} shows the ground state density of states (DOS)
for various interaction strengths
for the three regimes of superexchange interaction
of our phase-diagram.
We can see that for 
$U < U_c$, the DOS has a finite weight at the Fermi energy,
and for $U \geq U_c$, the DOS has a gap in the spectrum. 
As $U \rightarrow U_c$, the DOS develops a prominent dip 
at the Fermi energy, a signature of the pseudogap (PG) phase.
We can understand this in the following way.
The band ($U=0$) limit of this model is a metal, 
with finite DOS and a peak at the Fermi level.
Inclusion of the inter-orbital interaction ($U$) leads to
the emergence of orbital moments ${\bf \Gamma}_i$, with
the size of the orbital moment $|{\bf \Gamma}_i|$ determined by 
the strength of $U$. 
For $U < U_c$, we have $|{\bf \Gamma}_i| \ll \Gamma_{sat} = 1$.
(see fig.\ref{misc_T0}, discussed later)
The presence of these orbital moments reduce the DOS at the Fermi level.
As $U \rightarrow U_c$, $|{\bf \Gamma}_i|$ increases monotonically 
and for $U \gg U_c$ it saturates to the atomic value
$|{\bf \Gamma}_i| = 1$. The presence of large orbital moments
for $U \geq U_c$ leads to the opening of a gap in the DOS.
From our calculation, we estimate that
for $J_{AF}=0$, $U_c=11.0t$,
for $J_{AF}=0.2t$, $U_c=7.6t$ and 
for $J_{AF}=0.6t$, $U_c=5.0t$. 
The superexchange interaction favors
the Mott-insulating phase.

The lower set of panels in Fig.\ref{dos} show the DOS near the MIT for fixed
ratios of $U/U_c(J_{AF})$. Within each panel the $J_{AF}$ is varied to
probe if the spectral behaviour changes with changing AF coupling,
after factoring out the effect of $U_c$ change by normalising the
frequency axis by $U_c$. Our primary observation is that increasing
$J_{AF}$ leads to enhanced low energy DOS for a fixed ratio $U/U_c$.
We attribute this to the increased spin and orbital disorder in the
larger $J_{AF}$ situation - leading to an increasing `Anderson-Mott'
character of the metal-insulator transition. 
We have computed the inverse participation ratio (IPR) for states
as $J_{AF}$ is increased and find increasing localization. We
discuss those results later.

\begin{figure*}[t]
\centerline{
\includegraphics[angle=0,width=16.2cm,height=8.4cm]{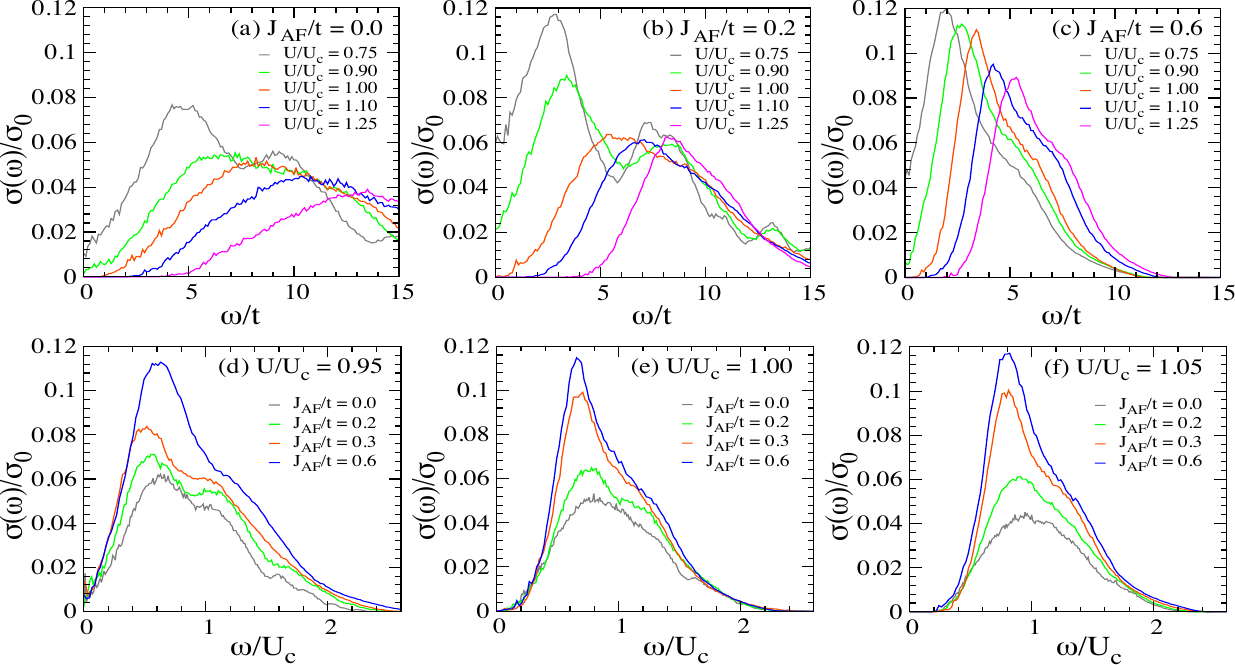}
}
\caption{\label{optics}
(a)-(c) Ground state optical conductivity for $J_{AF} =$ 0, 0.2t and 0.6t
for different $U/U_c$.
(d)-(f) Ground state optical conductivity for $U/U_c =$ 0.95, 1.0 and 1.05,
on a normalised frequency scale, for the indicated $J_{AF}$ values. 
}
\end{figure*}

\subsection{Optics and transport}

Fig.\ref{optics} shows the optical conductivity,
$\sigma(\omega)$, in the ground state for various interaction strengths 
and three regimes of superexchange interaction of our phase-diagram.

The band ($U=0$) limit of the model has finite DOS at the Fermi level.
As a result $\sigma(\omega)$ shows a Drude peak in this limit.
Inter-orbital interaction ($U$) leads to the emergence of 
orbital moments ${\bf \Gamma}_i$. 
For $U < U_c$, we have $|{\bf \Gamma}_i| \ll \Gamma_{sat} = 1$.
Increasing size of these orbital moments leads to a suppressed 
Drude response, and $\sigma(\omega)$ peak shifts to 
higher frequencies, indicating an increase in the insulating 
tendency of the system.

$|{\bf \Gamma}_i|$ increases monotonically with increasing $U$ 
and for $U \gg U_c$ it saturates to the atomic value
$|{\bf \Gamma}_i| = 1$. Beyond $U_c$ 
there is an optical gap in $\sigma(\omega)$.
From our calculation, we find that the $U_c$'s
for different superexchange scales are consistent with those
obtained from the DOS results.

\begin{figure}[t]
\centerline{
\includegraphics[angle=0,width=4.1cm,height=5.0cm]{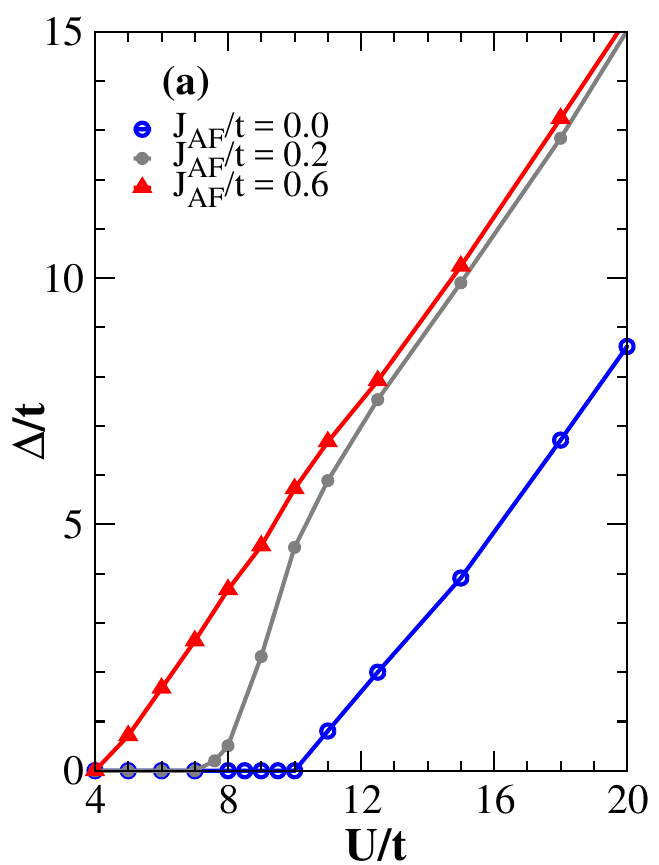}
\includegraphics[angle=0,width=4.2cm,height=5.0cm]{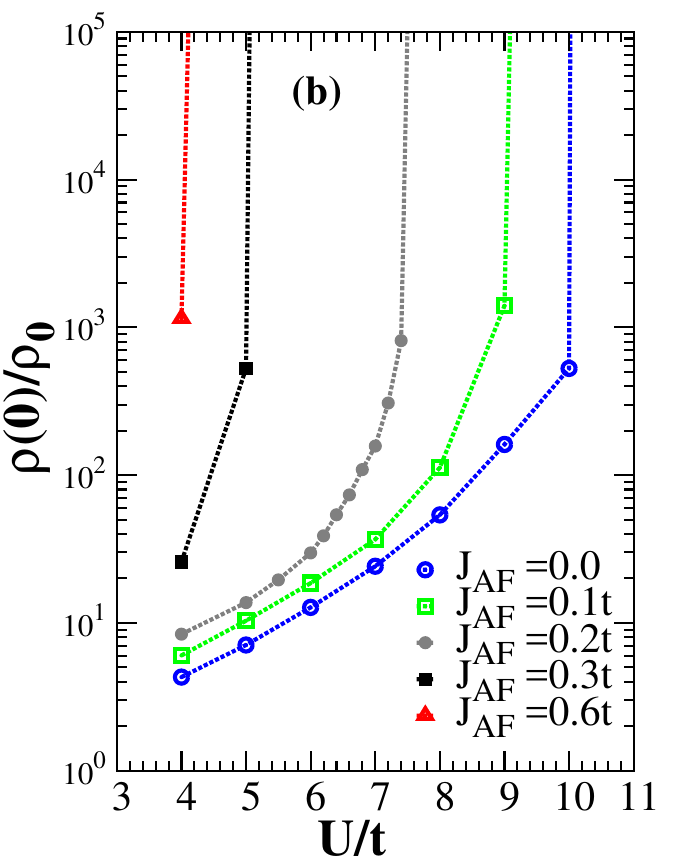}
}
\caption{\label{gap_res} 
(a)~Variation of optical gap ($\Delta/t$) with $U/t$ 
for different $J_{AF}/t$ values.
Panel (b) shows the variation of residual resistivity $\rho(T=0)$
with $U/t$ for different $J_{AF}/t$ values. The normalizing scale
is $\rho_0 = \hbar/e^2$.
}
\end{figure}

The lower set of panels in Fig.\ref{optics} 
show the optical conductivity near the MIT 
for fixed ratios of $U/U_c(J_{AF})$. 
Within each panel the $J_{AF}$ is varied to
probe if $\sigma(\omega)$ changes with changing AF coupling,
after factoring out the effect of $U_c$ change by normalising the
frequency axis by $U_c$. Our primary observation is the 
increase in the low frequency spectral weight at a fixed $U/U_c$
as $J_{AF}$, and the associated background disorder, increases.

We show the optical gap $\Delta$ in Fig.\ref{gap_res}(a).
It is clearly seen that $\Delta =0$ for $U < U_c$ and it increases 
monotonically for $U \geq U_c$.
Fig.\ref{gap_res}(b) shows the variation of residual dc resistivity,
$\rho(T=0)$ with $U/t$ for different superexchange values.
The finite $\rho(0)$ for $U < U_c$ can be understood by the scattering
of electrons from the (small) orbital moments. For $U \geq U_c$,
the (large) orbital moments lead to an opening of a Mott-gap
which manifests as $\rho(0) \rightarrow \infty$. 
These behaviors are seen in figure \ref{gap_res}.

\begin{figure}[b]
\centerline{
\includegraphics[angle=0,width=6.0cm,height=5.0cm]{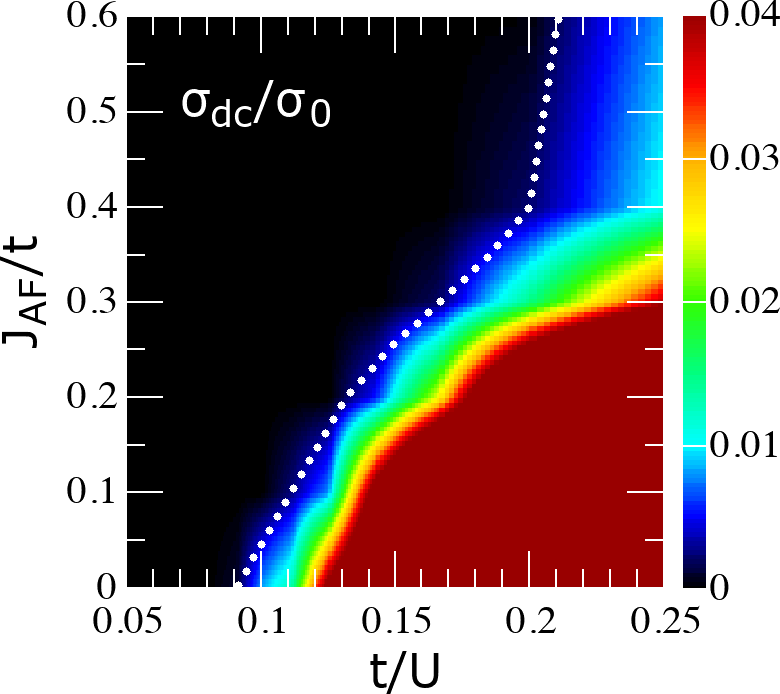}
}
\caption{\label{sigma_dc} Ground state d.c. conductivity, $\sigma_{dc}$,
for varying $t/U$ and $J_{AF}/t$. The normalizing scale is $\sigma_0 = e^2/\hbar$.
The MIT boundary can be thought of as the vanishing of $\sigma_{dc}$, 
with increasing $U/t$ values.}
\end{figure}

Fig.\ref{sigma_dc} shows the dc conductivity $\sigma_{dc}$ 
from our calculation in the $t/U$-$J_{AF}/t$ plane.  
We observe $\sigma_{dc}$ vanishing as, $U \geq U_c$.
This also allows us to roughly estimate the MIT boundary.

\begin{figure}[t]
\centerline{
\includegraphics[angle=0,width=4.0cm,height=4.0cm]{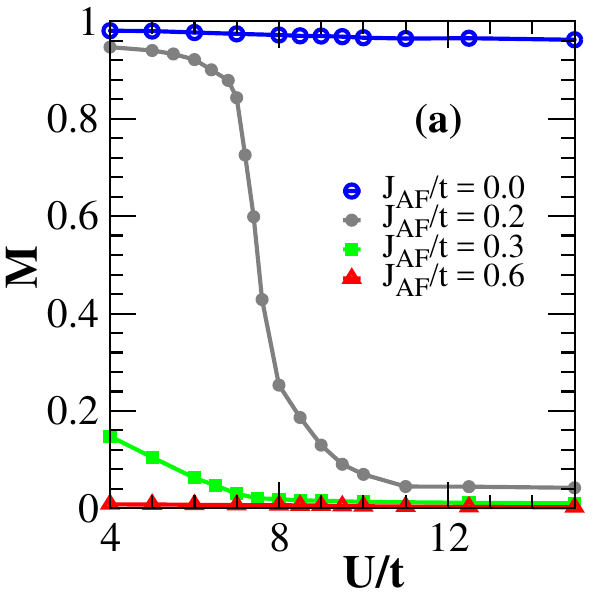}
\includegraphics[angle=0,width=4.0cm,height=4.0cm]{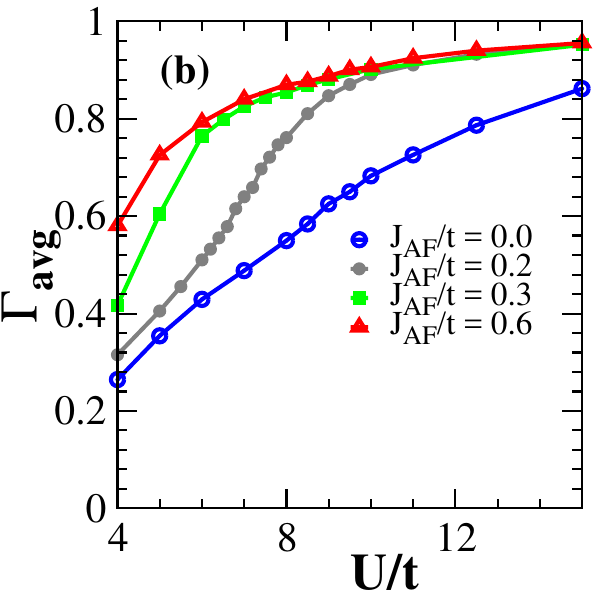}
}
~~\centerline{
\includegraphics[angle=0,width=4.1cm,height=4.0cm]{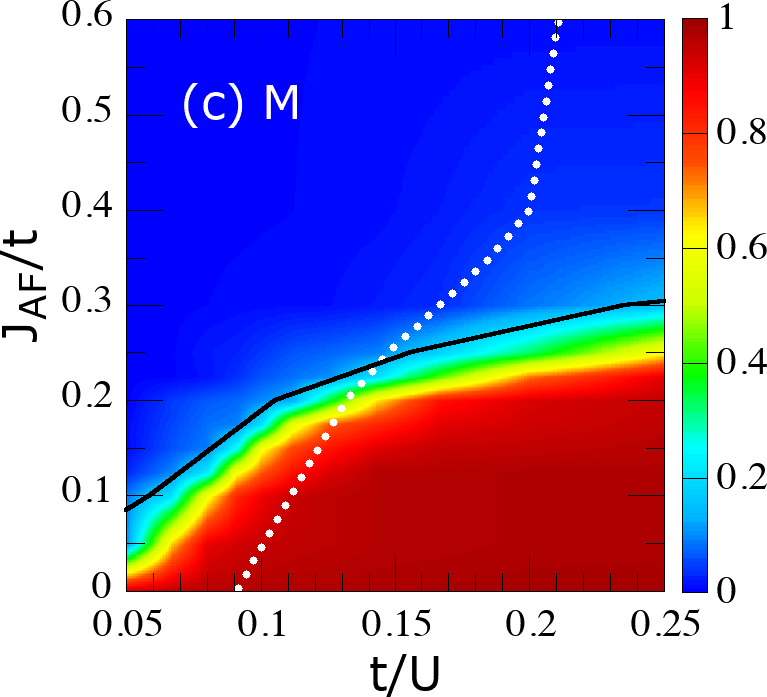}
\includegraphics[angle=0,width=4.1cm,height=4.0cm]{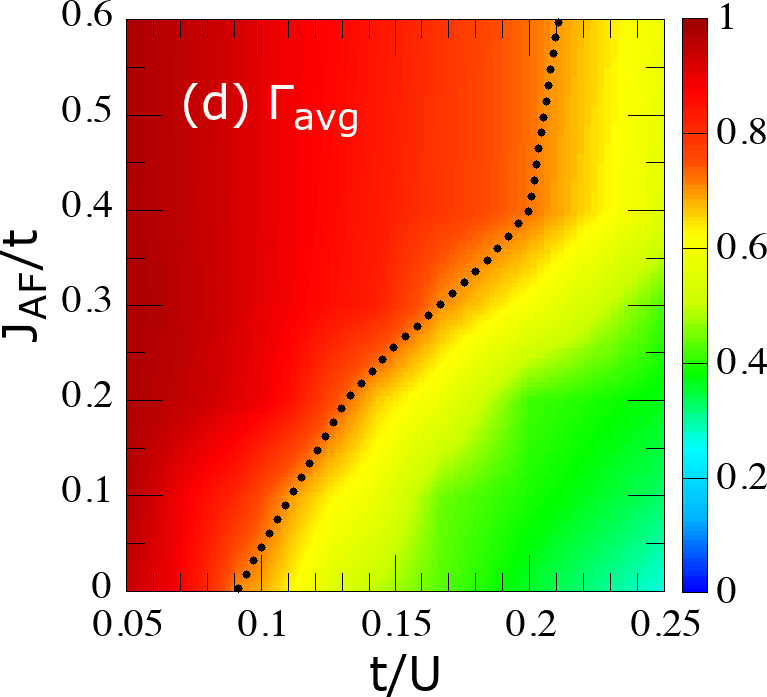}
}
\caption{
\label{misc_T0} Magnetisation $(M)$ and average orbital moment
$(\Gamma_{avg})$ in the ground state.
(a)~$U/t$ dependence of $M$ at $J_{AF} =0$ indicates that
the system has $M=1$ at all $U$, irrespective of metal/insulator
character, and for $J_{AF} \gtrsim 0.5t$, $M \sim 0$ for the entire
$U$ window probed. At intermediate $J_{AF}$, $M$ shows
a rapid crossover around a scale $U_{mag}(J_{AF})$ that
is close to but not quite the metal-insulator transition point $U_c(J_{AF})$. 
(b)~For $U/t \rightarrow \infty$, the orbital moment, $\Gamma_{avg} \rightarrow 1$, 
as expected in the atomic limit. The approach to this asymptote
is faster at larger $J_{AF}$. The $U \rightarrow 0$ behaviour
is dictated by the bandstructure, and change in the magnetic state with $J_{AF}$.
(c)-(d) Overall variation  of $M$ and $\Gamma_{avg}$ 
in the $J_{AF}/t$ and $t/U$ plane. The dashed line is the
MIT boundary separating the metallic and insulating regimes.
}
\end{figure}

In Fig.\ref{misc_T0}(a), we highlight the behavior of the magnetisation $(M)$ 
in the ground state. We find that at $J_{AF} =0$ the system
 has saturated magnetisation, 
($M=1$) at all $U$ values, irrespective of the metallic or insulating character. 
On the other hand, for $J_{AF} \gtrsim 0.5t$, the magnetisation is vanishingly small, 
($M \sim 0$) for the entire $U$ range probed in our study. For
 intermediate $J_{AF}$ values, 
the magnetisation displays a rapid crossover around a scale $U_{mag}(J_{AF})$ 
that is close to but not quite the metal-insulator transition point $U_c(J_{AF})$.
This is an indication of distinct energy scales governing the magnetic transition and 
the Mott transition in our model. 

In Fig.\ref{misc_T0}(b), we show the behavior of average orbital moment
$\Gamma_{avg} = {1/N}\sum_i \vert {\bf \Gamma}_i \vert$ in the ground state. 
For $U/t \rightarrow \infty$, the orbital moment $\rightarrow 1$, as
one expects in the atomic limit. The approach to this asymptote
is faster at larger $J_{AF}$ values. On the other hand, the $U \rightarrow 0$ behaviour
is dictated by the electronic bandstructure, and change in the magnetic
state with $J_{AF}$.
In Fig.\ref{misc_T0}(c)-(d), we show the overall variation of $M$ and $\Gamma_{avg}$ 
in the ground state, in the $J_{AF}/t$ and $t/U$ plane. It can be seen that 
the boundary separating the ferromagnetic and spin-liquid phases, and
the MIT boundary separating the metallic and insulating regimes are distinct ones.
We discuss the detailed nature of this interplay next.

\section{Discussion}

To get a feel for the changing magnetic state and 
the shifting MI transition point, it is useful 
to examine an approximate effective `spin only' model.
Consider the bond kinetic energy in a spin configuration
$\{ {\bf S}_i \}$. It is the product of
an electronic average and a 
modulated hopping both of which depend on 
$\{ {\bf S}_i \}$. The dependence of the spin
overlap  factor 
is explicit, it is simply:
$\sqrt{ (1 + {\bf S}_i.{\bf S}_j)/2}$. The electronic
average does not have an obvious expression in terms
of the spins but, as a starting approximation, we can
replace $ \langle c^{\dagger}_{i\alpha}c_{j\beta} \rangle$
by its thermal average \cite{SCR_DE}. The thermal average,
please note, is not a spin configuration dependent quantity.

\begin{figure}[b]
\centerline{
\includegraphics[angle=0,width=6.0cm,height=5.0cm]{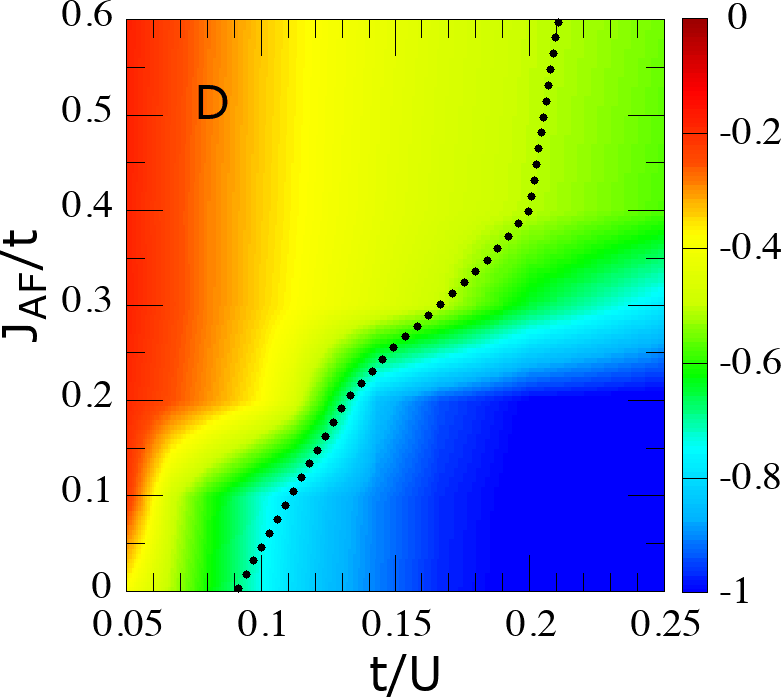}
}
\caption{\label{KE} The effective ferromagnetic exchange, $D$, at $T=0$
for varying $t/U$ and $J_{AF}/t$. The calculation and significance of
this quantity is explained in the text. The MIT boundary is shown
by dotted lines and coincides with change from large to small values of
$D$.}
\end{figure}

Under this assumption the kinetic energy term can be approximated as
below, and added to the AF term.
\begin{eqnarray}
H_{eff}\{ {\bf S } \} & \approx & 
\sum_{ij} D_{ij} \sqrt{ (1 + {\bf S}_i.{\bf S}_j)/2}
+ J_{AF} \sum_{\langle ij \rangle} {\bf S}_i.{\bf S}_j
\cr
D_{ij} & = &
\sum_{\alpha \beta} t_{ij}^{\alpha\beta}
\langle c^{\dagger}_{i\alpha}c_{j\beta} + h.c \rangle
\nonumber
\end{eqnarray}
The role of the Hubbard interaction, acting through the
orbital moment, is implicit in the model above. The 
$D_{ij}$ are supposed to be computed in backgrounds 
that include the ${\bf \Gamma}_i$ as well as the AF 
coupling.
Since the dependence of $D_{ij}$ on the magnetic and 
orbital state is not known the model above does not
have much predictive value. However, the thermally (and
system) averaged $D_{ij}$, which we call just $D$,
can serve to identify the origin of the changing magnetic character
(see figure~\ref{KE}).
It can also be related to direct measurables, {\it e.g},
(i)~the spin stiffness (spin wave velocity), since the
$D$ and $J_{AF}$ dictate this quantity, and (ii)~the
integrated optical weight, via the $f-$sum rule
$$
\sum_{ij} D_{ij} \sqrt{ (1 + {\bf S}_i.{\bf S}_j)/2}
~\propto~ \int_0^{\infty} \sigma(\omega) d \omega
\equiv n_{eff}
$$
where $n_{eff}$, the integrated
optical weight, is related to the effective carrier
density.
This can be roughly simplified to $D \sqrt{ 1 + m^2} 
\propto n_{eff}$, where we have approximated the spin
average by $m^2$.  
The physics content of this is simple -
reducing magnetisation reduces the 
hopping $(D)$ and the combination determines
$n_{eff}$.

\subsubsection{The metal-insulator transition line}

The role of $J_{AF}$ is to generate magnetic phase
competition and reduce the ferromagnetic tendency by 
suppressing the kinetic energy. 
To set a convenient reference, the effective bond
resolved kinetic energy, $D$,
at $J_{AF}=0$ and $U \rightarrow 0$
is $\sim -t$.
That allows us to set up three regimes.

(a).~When $J_{AF} \ll D$, we essentially have a weakly
renormalised FM ground state and $U_c$ is only
modestly suppressed with respect to the $J_{AF} =0$
value. For us this happens when $J_{AF} \lesssim 0.1t$.
(b).~In the interval $0.1t < J_{AF} < 0.4t$ 
the $U_c$ changes quickly, at $J_{AF} =0.4t$ it is roughly
half the value at $J_{AF}=0$.
(c).~For $J_{AF} \gtrsim 0.4t$ the $U_c$ does not reduce any further 
since the magnetic ground state is completely
disordered and the magnetisation cannot be
suppressed any further. This shows up as the vertical
asymptote of the MIT line in Fig. \ref{pd_T0}.

\subsubsection{The ferromagnet to `spin liquid' transition}

The ferromagnet to spin liquid `transition' 
occurs along a line that we call $U_{mag}(J_{AF})$. 
There is some ambiguity in locating this line 
since within our parameter space 
the magnetisation is always finite, if small.
We set $M = 0.05$ as the S-F to S-L transition. 
Just as $U_c$ is dictated roughly by the
competition between $U$ and $D$, $U_{mag}$ is
decided by the competition between $J_{AF}$ and $D$.

\subsubsection{Anderson character of the Mott insulating phase}

Increasing superexchange leads to disordered spin and orbital backgrounds.
As shown in Fig.\ref{dos} (e) and \ref{optics} (e), the optical gap survives 
even when the single particle spectral gap closes via spectral weight accumulation at the Fermi level. 
Furthermore, a closer examination of the states near the fermi-energy in this regime 
shows that the low-energy states are localized, as characterised by 
significantly large values of inverse participation ratio (see appendix). 
These localized states are responsible for a Mott insulating phase with finite optical gap, but no spectral gap. 
We attribute this behaviour to the onset of the ``Anderson-like'' character of the Mott insulating phase. 
As expected this behaviour is prominent for $J_{AF} \neq 0$, where the geometry of the pyrochlore lattice 
plays a crucial in the localization of states. With further increasing $U > U_c$, a gap appears in the density of states.
However, the optical gap continues to remain larger than the spectral gap, and the localized states still survive near the gap edges.
Thus the Mott insulating phase in our model is essentially of Anderson-Mott character.

\begin{figure*}
\centerline{
\includegraphics[angle=0,width=14.0cm,height=7.5cm]{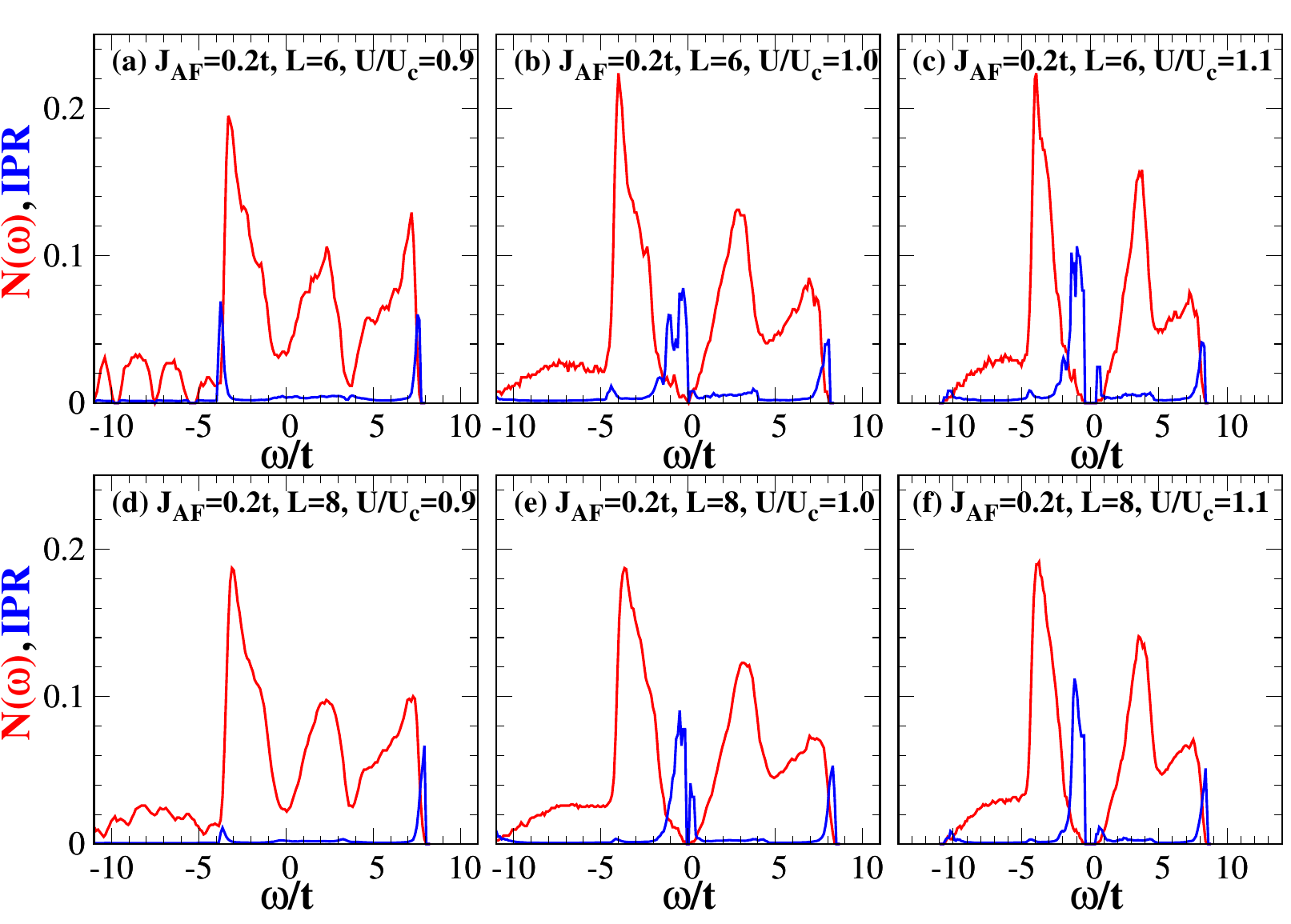}
}
\caption{\label{fig10} Single particle DOS ($N(\omega)$) and 
inverse participation ratio (IPR) at $T=0$ for system sizes $L=6$ ((a)-(c)) 
and $L=8$ ((d)-(f)), for a selected choice of $J_{AF} = 0.2t$ for 
$U=0.9U_{c}, U_{c}$ and $1.1U_{c}$, respectively.}
\end{figure*}

\section{Conclusion}

In this paper we have studied the ground state properties 
of two orbital Hubbard model 
with the electrons additionally strongly coupled to a background local
moment - and the moments interacting antiferromagnetically amongst
themselves. This Hubbard-double exchange-superexchange scenario,
on the pyrochlore lattice, is the minimal model for the rare earth
molybdates. We map out the ground state phase diagram via a 
simulated annealing based unrestricted Hartree-Fock calculation
and establish the metal-insulator and ferromagnet-spin liquid
transition boundaries.
We provide the detailed structure factors, the density of
states across the metal-insulator transition, and the
optical conductivity, pointing out an
increasing Anderson character to the notional `Mott'
transition as the
antiferromagnetic superexchange is increased. This effect should
be readily observable in the high pressure experiments.

\section{Acknowledgment}
We acknowledge use of the High Performance Computing cluster at 
Harish Chandra Research Institute, Prayagraj (Allahabad), India.

\section{Appendix}

Our measurement of spectral and optical gaps at the Fermi level reveals 
that unlike the naive expectation of them being equal, in the 
present system the magnitude of the optical gap exceeds that of 
the spectral gap. The observation can be attributed to ``disorder 
free" Anderson-Mott localization, wherein the geometric frustration 
of the lattice gives rise to localization of the electrons.
The corresponding single particle DOS exhibits accumulation of 
spectral weight at the gap edges, thereby reducing the spectral 
gap. 

In order to ascertain the localizing tendency of the moments we 
have calculated the inverse participation ratio (IPR) defined 
as, 
\begin{eqnarray}
IPR &=& \sum_{i, \alpha, \sigma}\vert u^{i}_{\alpha, \sigma}\vert^{4}
\end{eqnarray}
where, $u^{i}_{\alpha, \sigma}$ is the eigenvector corresponding to 
the eigenvalue $\epsilon_{\alpha}$. For an eigenstate $\psi_{i,\alpha}$ 
the localization length ($\xi_{loc}$) is related to the IPR as, 
$IPR \propto 1/\xi_{loc}^{2}$. Thus, an increase in localization leads 
to reduced $\xi_{loc}$ and increase in IPR. In Figure \ref{fig10} we 
show the single particle DOS along with the corresponding IPR for a 
selected $J_{AF} = 0.2t$, as a function of increasing $U/t$ (normalized 
with respect to 
$U_{c}$,  corresponding to the MIT at $J_{AF}=0$). Further, we have 
compared our results at two different system sizes which shows that 
our observations are robust against finite system size effects. Note 
that access to still larger system sizes is restricted by the computational 
expense. 

Based on Figure \ref{fig10} we infer that for $U \gtrsim U_{c}$, the  
opening of the Mott gap at the Fermi level is accompanied by a progressive 
increase in IPR close to the gap edges. We quantify the observed localization 
in terms of the fraction of localized states, which is the ratio between 
the number of localized states and the total number of states. 
As a function of increasing $U/U_{c}$ the fraction of
localized states for $L=6$, varies as, 0.165, 0.174 and 0.211
for $U=0.9U_{c}$, $U_{c}$ and $1.1U_{c}$, respectively.
For $L=8$, the fraction of localized states changes to, 
0.154, 0.179 and 0.217 for $U=0.9U_{c}$, $U_{c}$ and $1.1U_{c}$, respectively.
This indicates that $U/t$ favors the localizing tendency of the moments.

Our result suggests the possibility of disorder free localization of single 
particle eigenstates aided by geometric frustration of the underlying 
lattice. It must however be noted that our current numerical framework 
doesn't take into account the effect of quantum fluctuations, which can 
lead to dephasing of the single particle states. An analysis of the effect 
of quantum fluctuations on such frustration aided single particle localization 
is beyond the scope of this work. It is however expected that signatures of 
such localized states will survive at finite temperatures as well and will 
provide a possible mechanism to resolve the experimentally observed inequality 
between the spectral and optical gaps in geometrically frustrated pyrochlore 
lattices undergoing Mott transition. 

\bibliographystyle{unsrt}

\end{document}